\documentclass[runningheads,a4paper]{llncs}

\usepackage[english]{babel}

\usepackage[%
rm={oldstyle=false,proportional=true},%
sf={oldstyle=false,proportional=true},%
tt={oldstyle=false,proportional=true,variable=true},%
qt=false%
]{cfr-lm}
\usepackage{graphicx}

\usepackage{paralist}
\usepackage{graphicx}
\usepackage{subcaption}

\usepackage{cite}

\usepackage[T1]{fontenc}

\usepackage[math]{blindtext}

\usepackage{csquotes}

\usepackage{microtype}

\usepackage{url}
\makeatletter
\g@addto@macro{\UrlBreaks}{\UrlOrds}
\makeatother

\usepackage[table,xcdraw]{xcolor}

\usepackage[
bookmarks=false,
breaklinks=true,
colorlinks=true,
linkcolor=black,
citecolor=black,
urlcolor=black,
pdfpagelayout=SinglePage,
pdfstartview=Fit
]{hyperref}
\usepackage[all]{hypcap}

\usepackage{todonotes}

\usepackage{xspace}

\DeclareFontFamily{U}{MnSymbolC}{}
\DeclareSymbolFont{MnSyC}{U}{MnSymbolC}{m}{n}
\DeclareFontShape{U}{MnSymbolC}{m}{n}{
    <-6>  MnSymbolC5
   <6-7>  MnSymbolC6
   <7-8>  MnSymbolC7
   <8-9>  MnSymbolC8
   <9-10> MnSymbolC9
  <10-12> MnSymbolC10
  <12->   MnSymbolC12%
}{}
\DeclareMathSymbol{\powerset}{\mathord}{MnSyC}{180}

\usepackage{soul}

\usepackage[acronym]{glossaries} 
\newacronym{ast}{AST}{Abstract Syntax Tree}
\newacronym{ct}{CT}{Combinatorial Testing}
\newacronym{czt}{CZT}{Community Z Tools}
\newacronym{dap}{DAP}{Debug Adapter Protocol}
\newacronym{dbgp}{DBGP}{Common DeBugGer Protocol}
\newacronym{gui}{GUI}{Graphical User Interface}
\newacronym{ide}{IDE}{Integrated Development Environment}
\newacronym{lsp}{LSP}{Language Server Protocol}
\newacronym{poc}{PoC}{Proof of Concept}
\newacronym{pog}{POG}{Proof Obligation Generation}
\newacronym{po}{PO}{Proof Obligation}
\newacronym{slsp}{SLSP}{Specification Language Server Protocol}
\newacronym{vdm}{VDM}{Vienna Development Method}
\newacronym{vscode}{VS Code}{Visual Studio Code}
\newacronym{CSV}{CSV}{Comma Separated Values}

\usepackage{chngcntr}

\usepackage[capitalise,nameinlink]{cleveref}
\crefname{section}{Sect.}{Sect.}
\Crefname{section}{Section}{Sections}
\crefname{lstlisting}{Listing}{Listing}
\Crefname{lstlisting}{Listing}{Listing}

\usepackage{listings} 
\usepackage{times}    

\lstdefinelanguage{json}{
    basicstyle=\ttfamily\small, 
    numbers=left,
    stepnumber=1,
    numbersep=8pt,
    breaklines=true,
    frame=single,
    xleftmargin=.11\textwidth, 
    xrightmargin=.11\textwidth
}

\usepackage{vdmlisting}

\begin{document}
\pdfgentounicode=1

\title{Specification-based CSV Support in VDM}

\author{Leo Freitas\inst{1} \and Aaron John Buhagiar\inst{2}
}
\authorrunning{ }

\institute{School of Computing, Newcastle University, \\
\email {leo.freitas@newcastle.ac.uk}\\
\and Translational and Clinical Research Institute, Newcastle University\\
\email{a.j.buhagiar2@newcastle.ac.uk}
}
			
\maketitle
\setcounter{footnote}{0}
\begin{abstract}
CSV is a widely used format for data representing systems control, information exchange and processing, logging, \textit{etc}. Nevertheless, the format is riddled with tricky corner cases and inconsistencies, which can make input data unreliable, thus, rendering modelling or simulation experiments unusable or unsafe. We address this problem by providing a \textbf{SAFE}-CSV VDM-library that is: \textbf{S}imple, \textbf{A}ccurate, \textbf{F}ast, and \textbf{E}ffective. It extends an ecosystem of other VDM mathematical toolkit extensions, which also includes a translation and proof environment for VDM in Isabelle. 
\end{abstract}


\section{Introduction}\label{sec:intro}

The \gls{CSV} format is widely used for a variety of applications:~from data science as in organ transplant allocation\footnote{\url{https://unos.org/data/}}, embedded systems representation as in state machines for medical devices~\cite{alastairMSc,scp-dialyser}, pharmaceutical applications~\cite{csv-validation-pharma-msc}, log files on various kinds of systems, databases, payments, government systems and so on. Despite there being a standard (RFC 4180)\footnote{\url{https://www.rfc-editor.org/rfc/rfc4180}}, there are many versions and variations\footnote{\url{https://commons.apache.org/proper/commons-csv/}}. Its long history of use\footnote{\url{https://bytescout.com/blog/csv-format-history-advantages.html}} has inspired various other ``simple'' formats for data management and exchange, such as JSON~\cite{JSON}, XML and various spreadsheet formats. 

CSV processing is an important part of the computation involving multiple application domains. Areas dependant on CSV input are plagued with subtle errors and inconsistencies, which would ultimately invalidate the systems responses associated with CSV data. When such data represents a system's architecture in itself (\textit{e.g.}~dialyser state machine representation~\cite{scp-dialyser}), as opposed to system data, consequences can be catastrophic. 

To be clear, we delegate CSV parsing (\textit{i.e.}~handling of CSV grammar/format itself) to external tools. Here we present a VDM CSV library to capture aspects of the CSV data being imported to VDM, such that users can introduce specification to the CSV data as if it was a VDM data type. For system-architecture CSV files, tools like VDMJ's \texttt{CSVReader} or bespoke tools like those in~\cite{egleUG,alastairMSc,emv2} are suited; whereas for general data manipuation over varied CSV dialects and IO speed (\textit{e.g.}~log files, simulation runs, \textit{etc.}), our proposed \texttt{CSVLib} library is better suited.

Therefore, we argue it is important to have a formal characterisation of the data structures represented by CSV input, with enough flexibility and variation to cater for realistic applications. For that, we choose VDM.\@ For formal characterisation of the CSV-file format itself, we delegate this to formal grammars\footnote{For an example formal CSV-grammar, see~\url{https://github.com/antlr/grammars-v4/tree/master/csv}} and parsing tools.\@ 

The \gls{vdm} has has been widely used both in industrial contexts and academic ones covering several domains, such as Security~\cite{Kulik&20,Kulik&21a}, Fault-Tolerance~\cite{Nilsson&18}, Medical Devices~\cite{Macedo&08}, among others. We extend VDM specification support with a suite of tools and mathematical libraries\footnote{\url{https://github.com/leouk/VDM_Toolkit/}}. The work is also integrated within the VDM \gls{vscode} IDE.\@

We decided to depart from the standard VDM-CSV library because it is slow, cannot handle various CSV dialects, relies on CSV positioning management (\textit{e.g.}~knowing where in the file you are; reading one CSV line at a time), and has no checks on the CSV data itself.        

In this paper we report on the recent extension of the VDM toolkit to support CSV format parsing, validation and printing. We illustrate its use with a variety of scenarios frequently seen in practical uses of CSV.       

\section{Background}~\label{sec:background}

CSV is plagued with a variety of fiendish scenarios leading to unexpected errors~\cite{csv-validation-pharma-msc,10.1145/3274856.3274879,alastairMSc}. For critical applications relying on the format, this is particularly problematic. This motivated the creation of a formal environment to capture CSV problems clearly and concisely, and to validate outcomes according to user-defined invariants of different nature, for example, data type invariants on CSV cells and consistency invariants on row or column data, such as column ordering or row data redundancy consistency (\textit{e.g.}~weight, height and BMI info must be correlated). Finally, also an overall file invariant. Furthermore, CSV files can often contain implicit defaults or inconsistent correspondences. Thus, we are interested in capturing such constraints formally using a VDM library (\texttt{CSVLib}). We integrate this library within the VDM VSCode extension\footnote{\url{https://github.com/overturetool/vdm-vscode}}, as well as VDMJ~\cite{Battle09}. 

Related to formal CSV (and data sources) formal processing, we worked on generation of VDM files from XSD files describing the data dictionary of the EMV payment systems~\cite{emv2}, and similar efforts have been applied to the Function Mockup Unit standard~\footnote{\url{https://github.com/INTO-CPS-Association/FMI-VDM-Model}}. Another example is for loading the control system finite state machine definitions in CSV for a dialyser~\cite{egleUG} and a brain pace maker~\cite{alastairMSc}. These read in data in the corresponding (CSV or XML) format(s), and generate a bespoke VDM model corresponding to implicitly understood design decisions. 

These applications of CSV to VDM are different from our approach here, which focuses on raw CSV data loaded within a VDM data type that can then be further checked for various structural consistencies through dynamic invariants. Similar to our CSV library (\texttt{CSVLib.vdmsl}) is the one defined within VDMJ standard libraries (\texttt{CSV.vdmsl}): it provides users with native calls to a simple CSV parser, returning a sequence of wildcard (\texttt{?}) types. 


\section{Design Principles}\label{sec:principles}

Our VDM CSV library has four core design principles:

	\begin{enumerate}
		\item \textbf{S}imple:~ease of use is crucial, given CSV processing is a pervasive task;
		 
		\item \textbf{A}ccurate:~CSV input errors accounts for a considerable amount of modelling process inaccuracies; hence, we wanted a solution where multiple forms of expected validation were possible;
		 
		\item \textbf{F}ast:~CSV input can often be large; hence, varied and computationally efficient IO parsing solutions are important~\footnote{\url{https://github.com/uniVocity/csv-parsers-comparison}};
		 
		\item \textbf{E}ffective:~there are multiple CSV format-variants~\footnote{\url{https://commons.apache.org/proper/commons-csv/}}; hence, tolerance of format variations is important.   
	\end{enumerate}

These \textbf{SAFE} principles underpin the overall library design goals. Its architecture is divided in three parts:~i) native calls to IO;~ii) various CSV invariants checking mechanisms per cell, across row and across column, and for the overall file;~and iii)~a series of supporting functions and operations to enable easy access to CSV functionality. 

Moreover, we accept that real CSV applications are riddled with errors and inaccuracies. Thus, we also provide rich error reporting and support for CSV that betrays expected invariants to be processed and accurate information be given to users.  Once validation has taken place, users can confidently access the CSV data to their target end, and also print it out (after processing or otherwise) to a file. 


\section{Library architecture}\label{sec:architecture}

\begin{figure}[htbp]
   \centering
       \includegraphics[width=0.90\textwidth]{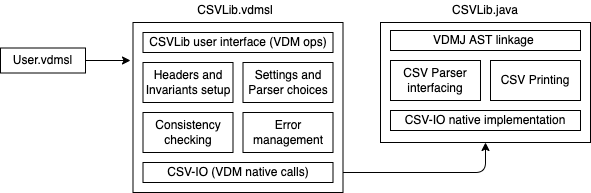}
   \caption{\texttt{CSVLib.vdmsl} architecture.}\label{fig:CSVLibArchitecture}
\end{figure}

Next, we present how the library architecture that implements our design principles (see~\Cref{fig:CSVLibArchitecture}). User models have to import \texttt{CSVLib.vdmsl}, which provides CSV-IO (\textit{e.g.,}~parsing and printing) as VDM native function calls. VDM native functions are a mechanism for linking VDM specification with an underlying Java implementation servicing each of the VDM native calls defined. For example, VDMJ provides native calls for some trigonometric functions, random value generation, and so on.  

\noindent\texttt{CSVLib.vdmsl} provides various functionalities including:
\begin{enumerate}
    \item Settings:~line comment, skipping blank lines, \textit{etc.};
    \item Parsers:~implementations geared for speed, multiple CSV formats, \textit{etc.};
    \item Headers:~strongly typed and with default values;
    \item Invariants:~checks over cells, across rows and columns, and across overall file; and
    \item Error handling:~cell-located errors with explanatory reason(s) for failure.
\end{enumerate} 

\subsection*{\texttt{CSVLib} use}

\texttt{CSVLib} has two entry points: either direct calls to CSV-IO native calls, or via operations on a state-based specification of expected library usage. The former provides the most flexible access to the CSV functionality, whereas the later provides the most convenient access. For instance, the simplest use of the CSV library is a call to the operation:
%
\begin{vdmsl}[frame=none,basicstyle=\ttfamily\scriptsize]
    loadSimpleHeadersCSV(path, <Native>, [<Str>,<Int>,<Bool>], false);
\end{vdmsl}
This will load the CSV file on \texttt{path} using the given parser interface implementation. We currently support two interfaces:~our own native implementation and \texttt{uniVocity} CSV, one of the fastest we could find\footnote{\url{https://github.com/uniVocity/csv-parsers-comparison}}. Other implementations are possible/planned. 

The list of quote values contain the types of the expected CSV columns, whilst the final argument states that processing will not be strict. That means, any failed invariant checks will be tolerated and recorded for further user inspection in the state. Otherwise, if strict was \texttt{true}, then invariant failures will not be tolerated and the result is an empty CSV\@.      

\subsection*{CSV headers}\label{subsec:Headers}

Many CSV files contain no header. Nevertheless, they ought to have a notion of type correctness for each value read (\textit{i.e.}~an implicit type). In \texttt{CSVLib}, headers must define a type and a default value. Defaults are important for common situations like comma-sequences in a row. Types are important to enable type consistency and invariant checking, whereas default values are important so that we can infer what was the hidden intention behind empty fields, and to ensure defaults satisfy type invariants. For instance, the header below states that age is an integer with \(18\) as its default value:
\begin{vdmsl}[frame=none,basicstyle=\ttfamily\scriptsize]
    mk_Header("Age", <Integer>, 18, cell_inv, col_inv)
\end{vdmsl}
\texttt{CSVLib} headers may also contain a description and two invariants:~one for every cell underneath a header, and another for the overall column.  

\subsection*{CSV invariants}\label{subsec:Invariants}

We have four CSV invariants, which are captured by the following VDM function types: 
\begin{vdmsl}[frame=none,basicstyle=\ttfamily\scriptsize]
    CSVCellInv = (CSVType * CSVValue -> Reason);
    CSVRowInv  = (Headers * Row -> Reason);
    CSVColInv  = (Header  * TransposedRow -> Reason);
    CSVFileInv = (Headers * Matrix -> Reason);
\end{vdmsl}
Invariant checks return a \texttt{Reason}, which is either \texttt{\textbf{nil}} in case the invariant is satisfied, or a non-empty string explaining the cause of the failure. This enables the specifier fine-grained control over errors occuring within a CSV file. Cell invariants receive the declared header type and the read cell value to be checked. Row invariants receive all readers and current row of cells to be checked. Column invariants receive the column header and the current column of cells to be checked, viewed as a transposed row. Finally, a file invariant defines properties across the whole CSV matrix.     

Some invariant kinds have an implicit check, regardless of whether a user-defined invariant is given or not. For cells, this checks the values against their corresponding declared header types, whereas for rows, we implicitly check that the header size corresponds to row size (\textit{e.g.}~no short rows); there are no implicit column or file invariant checks. 

For the \texttt{Age} header above, users could define \texttt{cell\_inv} (L1--3) to enforce age limits, whereas \texttt{col\_inv} (L5) could enforce no age uniqueness as:
\begin{vdmsl}[frame=none,basicstyle=\ttfamily\scriptsize,numbers=left,caption={Cell and Column Invariant Definitions},label={lst:CSVInvs}, numbers=none]
    (lambda -: CSVType, v: CSVValue & 
        if v < 18 then "below minimal age" else
        if v > 65 then "above maximal age" else nil);

    (lambda h: Header, c: TransposedRow & 
        if card elems c <> len c then "no duplicate ages are allowed" else nil);
\end{vdmsl}
%
Row invariants are useful for checking dependencies/redundancies across the row cells. If the CSV file had three extra \texttt{<Float>} headers for weight, height and BMI (Body Mass Index), there are interesting checks possible. For instance, given age type range (\texttt{\{18\ldots 65\}}) it is possible to presume some minimum height expected (\textit{i.e.}~row dependency invariant); and given BMI's formulae (\(\frac{height}{weight^2}\)), its cell value can be calculated (\textit{i.e.}~row redundancy invariant).    

File invariants acts upon on the whole dataset. This might be useful when certain properties needs to be correlated across multiple rows and columns. An example of this would be the importation of an end-of-year financial records, where individual taxes are calculated for each row and their summation is correlated to some previously established value. 

\subsection*{CSV VDM native calls}\label{subsec:VDMnative}

\texttt{CSVLib} has four native calls: three functions implementing file status, CSV read and write;~and one operation implementing low-level IO (not CSV) errors.   
\begin{vdmsl}[frame=none,basicstyle=\ttfamily\scriptsize]
    file_status   : Path -> FileStatus;
    csv_read_data : Path * CSVParser * CSVSettings * Headers -> 
                    bool * Errors * Data;
    csv_write_data: Path * Data -> bool;
    lastError     : () ==> Reason;
\end{vdmsl} 

\texttt{csv\_read\_data} implements the link with the Java CSV parsers. It expects a valid path, a known parser, settings and valid headers; it returns a tripple. In the result, if the success flag is \texttt{\textbf{false}}, some unexpected Java error occurred that can be inspected by a call to \texttt{lastError()}; the other two parts of the tripple representing short-row errors and data are empty. Otherwise, if the success flag is \texttt{\textbf{true}} and errors are not empty, then CSV parsing has identified short rows (\textit{e.g.}~row's size smaller than header's size), and they contain details where such errors occured in the CSV file. If the success flag is \texttt{\textbf{true}}, then the resulting data will have the CSV data loaded, yet without invariant checks. This allows loading invalid data, which can then be processed and errors reasons can be given to the user. Finally, if the success flag is \texttt{\textbf{true}} and errors are not empty, data will not contain the short rows identified.        

\subsection*{CSV Java native implementation}\label{subsec:JavaNative}

The implementation of VDM native calls is done in Java. The \texttt{CSVLib.java} file implements the VDM link, as well as various specialised VDMJ value-AST handling methods. These might be useful for other libraries requiring handling VDMJ values within VDM native method call implementations in Java. For example, we added for the value-AST in Java the VDM equivalent of record \texttt{mu}-expressions for VDM-record updates. This made the VDM record construction and update process in Java much like one would do in VDM itself. For instance, the \texttt{loadSimpleHeadersCSV} call above does not provide a header name. This will come from the CSV itself, and must update the header name field akin to \texttt{mu(header, name |-> "Age")}. In Java, this is performed as 
\begin{lstlisting}[language=Java,basicstyle=\footnotesize\ttfamily]
    namedHeaderAtI = ValueFactoryHelper.muRecord(headerAtI, 
        ValueFactoryHelper.mkFieldMap(
            Arrays.asList(HEADER_FIELD_NAME), 
            ValueFactoryHelper.mkValueList(
                ValueFactoryHelper.mkString(nameStr)), 
            Arrays.asList(true), ctx
        ), ctx);
\end{lstlisting}
\noindent The \texttt{ValueFactoryHelper} class provides a number of useful VDM value-AST support methods. The one above performs the record update on the field name given with the value passed. This illustrates how the \texttt{CSVLib} library can also be used by other libraries wanting to perform complex VDM value-AST manipulations. 

Finally, \texttt{CSVLib.java} delegates to a separate \texttt{CSVParser} Java interface, which provides the necessary services needed by the native call implementations. This is important in order to separate the VDMJ value-AST processing from the low-level CSV-IO parser implementations. This makes extending the library with other CSV parsers a relatively straightforward process:~users can implement the interface with their preferred CSV parser, and add the corresponding jar-file to the VDM-VSCode classpath. 
\begin{lstlisting}[language=Java,basicstyle=\footnotesize\ttfamily]
public interface CSVParser {
    public Iterator<String[]> parseCSV(final Reader reader) 
        throws IOException;
    public String lastError();
    public void clear();
    public CSVSettings getSettings(); } 
\end{lstlisting}
Any low-level implementation only has to return a Java iterator view of the input stream for the \texttt{CSVLib.java} to work according to the design principles described here. Other methods are self-explanatory. We use Java \texttt{Reader} as the preferred input type in order to take file encoding issues into account.  

The separation between Java implementation of VDM native calls and other Java code is important because debugging of VDM native calls alongside VDM specification is relatively tricky to setup. Such considerations and difficulties have now been resolved and can be reused by other VDM library developers that require VDM native call implementations. In practice, the setup enables the library developer to use VDMJ's command-line debugger to handle VDM library specification debugging, whilst using VSCode Java debugger to tackle Java's implementation of VDM native library calls. 
 
\subsection*{CSV data and errors}\label{subsec:DataAndErrors}

The CSV data type contains CSV settings, headers, and data matrix as a sequence of sequence of values. In \texttt{CSVLib}, an error occurs when one of the invariant check fails. A CSV error records cell position (\textit{i.e.},~row and column) alongside a non-empty explanation as to why the error has occurred.    
\begin{vdmsl}[frame=none,basicstyle=\ttfamily\scriptsize]
  Data :: settings: CSVSettings headers: Headers matrix: Matrix;
 Error :: rowNo: nat1 colNo: nat1 reason: Reason;
\end{vdmsl}  
\noindent After a successful call to \texttt{csv\_read\_data}, users can inspect what invariants have been violated by calling the function 
\begin{vdmsl}[frame=none,basicstyle=\ttfamily\scriptsize]
    csv_invariants_failed: Data -> set of Error
\end{vdmsl}
\noindent It returns the set of errors with the (row/col) position and a reason why the invariants have failed.  

\subsection*{CSV state and operations}\label{subsec:StateOps}

CSV VDM native calls and error handling are expressive and accessible through functions and types described above. Nevertheless, their expressivity can lead to involved specifications. To avoid this for end users, we provide a state-based interface with operations that give access to the CSV data matrix, any errors found, as well as other (IO) or usability errors.     
\begin{vdmsl}[frame=none,basicstyle=\ttfamily\scriptsize]
    state CSV of 
        file  : [Path]   parser: CSVParser       ferr: Reason
        strict: bool     pos   : set of Errors   data: Data
    inv mk_CSV(file, -, -, strict, pos, data) == 
       (file <> nil => file_status(file) = <Valid>) and
       (strict => csv_invariants_failed(data) = pos = {})...
\end{vdmsl}
The CSV state contains the CSV file path and parser kind, low-level (IO and other non-CSV) error reasons, strict invariant validation, positions of invariant errors, and CSV data matrix. We show some of the state invariants, which say that non-nil file status must be valid (\textit{e.g.~} exist, not be a directory, \textit{etc.}), and \texttt{strict} CSV does not tolerate any errors (including short row IO errors). The state and operations effectively illustrate how the VDM native calls can be used.     

There are three operations for end users. The \texttt{loadSimpleHeadersCSV} operation example above calls \texttt{loadCSV} with an appropriate headers created. The \texttt{loadCSV} operation ensures the file path is valid, (re-)sets up all relevant state parameters and loads the CSV data. Data loading follows the logic explained for VDM native call \texttt{csv\_read\_data}, where various cases are given descriptive error and/or information messages to the user. At this stage, both CSV data and invariant check errors are available as part of the state for inspection and further processing.  
\begin{vdmsl}[frame=none,basicstyle=\scriptsize\ttfamily]
    loadSimpleHeadersCSV: Path * CSVParser * seq1 of CSVType * bool ==> ();
    loadCSV       : Path * CSVParser * CSVSettings * Headers * bool ==> ();
    printCSV      : Path ==> ();
\end{vdmsl}

Finally, the operation \texttt{printCSV} prints out the loaded CSV to the given file path. This presumes loading has taken place successfully, and that the given path is not the same as the CSV file path itself to avoid data overwriting, and a corresponding CSV file will have been created (or overwrriten).  

\section{Evaluation}\label{sec:Evaluation}

The high-performance of our SAFE-CSV library is a core design principle. To demonstrate this, several CSV files, ranging from \(100\) to \(35,000\) rows were used to compare the performance of our library against the standard CSV library in VDM.

The standard CSV library provides limited features in terms of the variety of formats it can accept for the CSV input, its low-level IO mechanism (\textit{i.e.,}~read line by line) and the fact there are no data validation capabilities available. The scope of the performance testing was limited to largest subset of features allowed by the standard CSV library. SAFE-CSV provides a richer approach to data validation than VDM CSV;~however, these features were not used for this section to maintain parity between the two libraries during the performance test.

As SAFE-CSV also has the ability to change the underlying parser being used by the library, for this work, SAFE-CSV performance was tested using the Native parser and, the faster Univocity parser\footnote{\url{https://github.com/uniVocity/csv-parsers-comparison}}. 

\begin{figure}[ht]
    \centering
        \includegraphics[width=\textwidth]{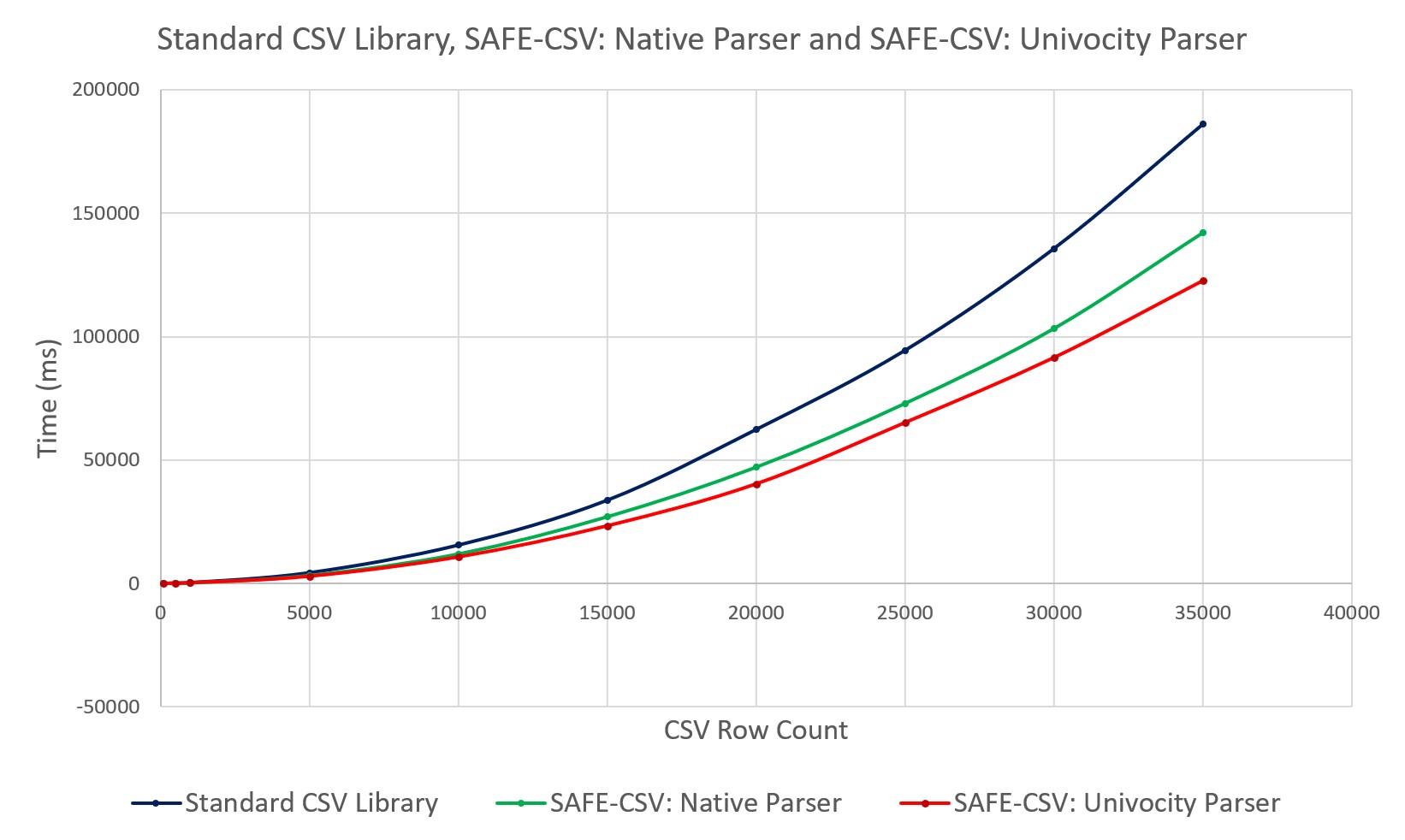}
    \caption{Standard CSV library and SAFE-CSV library (using the Native and Univocity parser) performance comparison}\label{fig:CSVPerformanceChart}
 \end{figure}

 Figure~\ref{fig:CSVPerformanceChart} shows the performance of the libraries using the same input data. The SAFE-CSV parsers performed significantly better than the standard library showing a speed-up of \(\sim 1.5\)x. The choice of underlying parser for the SAFE-CSV library has an effect on performance as the Univocity parser performed even better than the Native parser. With larger datasets, the Univocity parser also averaged a higher speed-up rate even compared to the Native parser.

\section{Examples}\label{sec:Examples}

Next, we illustrate the intended use of our library with a few examples. First, users have to provide information about CSV headers. In our example, we have a list of names with their corresponding age, weight (in kg), height (in cm), and BMI. We then create the corresponding VDM header as:
\begin{vdmsl}[frame=none,basicstyle=\ttfamily\scriptsize]
	EXAMPLE_HEADERS : Headers = 
        [mk_Header("Name"      , <String> ,  "Name", nil, COL_INV_UNIQUE_NAME),
         mk_Header("Age"       , <Integer>,       MIN_AGE, CELL_INV_AGE, nil),
         mk_Header("Weight(Kg)", <Float>  , MIN_WEIGHT_KG, CELL_INV_WEIGHT, nil),
         mk_Header("Height(cm)", <Float>  , MIN_HEIGHT_CM, CELL_INV_HEIGHT, nil),
         mk_Header("BMI"       , <Float>  ,       MIN_BMI, CELL_INV_BMI, nil) ];
\end{vdmsl} 
Default values are provided to test for CSV comma-sequences (\textit{e.g.}~CSV cells without any value), alongside cell and column invariant examples. These default values are used to ensure the implicit row invariant check that row size musts match header size. The cell invariants on age, weight, height and the column invariant on uniqueness are similar to the one shown above (see~\Cref{lst:CSVInvs}).

Next, we want to define a row invariant that checks the redundant BMI fields are consistent with respect to height and weight. Arguably the CSV should not have a BMI field. Having said that, many CSV files do contain such redundant information, which is frequently not accurate with respect to intended values. The BMI check row invariant is defined as:
\begin{vdmsl}[frame=none,basicstyle=\ttfamily\scriptsize]    
    (lambda h: Headers0, r: Row &
    	  if len h < 5 then "invalid BMI header"
    	  else 
		      (let bmi: real = calculated_bmi(r) in
			      if approx_eq(r(5), bmi, PRECISION) then nil 
			      else "invalid BMI for given CSV weight and height"));
\end{vdmsl}
For the given row, the user-defined function \texttt{calculate\_bmi} calculates the BMI using the row values for height and weight, whereas \texttt{r(5)} gets the CSV BMI value, which needs to be approximately equal to the calculated value. Approximation here considers a particular number of digits of precision for comparison between the cell value and the calculated one.    

Other examples are provided in \texttt{CSVLib.vdmsl} and \texttt{ CSVExample.vdmsl}\footnote{\url{https://github.com/leouk/VDM_Toolkit/tree/main/plugins/vdmlib}}. They show CSV parsing with different IO-parsers, CSV printing, escaped quotes in string cells (\textit{e.g.}~strings across multiple lines), default values (\textit{e.g.}~comma-sequences in CSV row), short rows (\textit{e.g.}~rows smaller than expected header), various types of invariant violation (\textit{e.g.}~implicit and user-defined), and so on.

\begin{lstlisting}[]
CSV (IO) error: ignoring 1 short rows from "CSVExample.csv"
    (1, 5): "CSV row 1 is too short for header: 
                expected 5 columns found 4 columns"
CSV invariants failed for "CSVExample.csv": 
                14CSV invariant failure at 14 cells: 
    ...

    (3, 2): "Invalid cell invariant: below minimal age"
    (4, 1): "Invalid col  invariant: repeated names"
    (4, 2): "Invalid cell invariant: above maximal age"
    (4, 3): "Invalid cell invariant: above maximal weight"
    (4, 5): "Invalid cell invariant: above maximal BMI"
    ...  
\end{lstlisting}

Output generated due to an import issue displays the violating cell in as a row/column tuple to aid the user in identifying and correcting the issue. Error output due to a custom invariant provides the user defined message as well the type of invariant it is (cell, row, column, file). Implicit failure such as short rows provided a standardised output.

\section{Results and discussion}\label{sec:Results}

In this paper, we presented a formally defined CSV library in VDM that adheres to our \textbf{SAFE} design principles (in~\Cref{sec:principles}). The library architecture (\Cref{sec:architecture}) is \textbf{S}imple, given its layered access to functionality and ultimately near-trivial user-interface access points. It is also \textbf{A}ccurate, given the presence of multiple kinds of user-defined invariants and other structural and data validation checks, such as detection of short rows and cell value type consistency with respect to declared headers. It is \textbf{F}ast, as the architecture layout allows for plug-and-play of different CSV-IO parsing. Finally, it is \textbf{E}ffective, given its combination of speed, ease of use, multiple capabilities around CSV format handling, CSV settings, errors handling, and so on.   

Compared to the standard VDM CSV library, our SAFE-CSV performed better, yielding over a \(50\)\% increase in speed. Whilst the more robust invariant system of the SAFE-CSV library was not used for performance analysis, the infrastructure for its execution and in-built type-checking are still present throughout. The performance impact of invariant checking is negligible. It is comparable to users encoding such check inv VDM themselves, on top of the standard VDM CSV library for IO anyhow.  

A key motivation for this work was the need for a more robust data import functionality to analyse data produced by an organ preservation medical device. The standard VDM CSV library proved to be too limited in the CSV format it can handle. Additionally, it provides no data validation capabilities, which are necessary for safety critical applications. 

\paragraph*{Future Work.}~
In future, we want to integrate the debugging environment of VDM native library calls, such that we use VDM VSCode DAP protocol~\cite{AdvancedVSCodePaper}. Moreover, on the CSV format itself, nested CSV formats, further CSV types, multiple CSV headers per file, \textit{etc}. This way, the VDM library developer can handle VSCode debugging of both VDM library specification and examples, alongside VSCode Java debugging. 

\subsubsection*{Acknowledgements}
We acknowledge NCSC funding on development of safer payment protocols. We very much appreciate fruitful discussion with Nick Battle about VDMJ native calls setup. 

\bibliographystyle{splncs03}
\bibliography{main.bib}


\end{document}